%% file: arXiv.Comment.CG2019.FrewerKhujadze.tex
\documentclass[11pt,a4,twoside]{article}
\usepackage[dvips,a4paper,left=2.5cm,right=2.5cm,top=2.5cm,bottom=2.5cm,headsep=1em]{geometry}
\usepackage{fancyhdr}
\usepackage{titlesec}
\usepackage[latin1]{inputenc}
\usepackage{amsmath,amsfonts,amssymb,array,amsthm}
\usepackage{rotating}
\usepackage[font={small}]{caption}
\usepackage{natbib}
\usepackage{lmodern,slantsc}
\usepackage{multirow}
\usepackage{chngcntr}
\usepackage{placeins}
\usepackage{caption}
\usepackage{enumitem}
\usepackage{maplestd2e}

\DefineParaStyle{Maple Output} \DefineParaStyle{Maple Output}
\DefineCharStyle{2D Math} \DefineCharStyle{2D Output}
\usepackage[breaklinks,
colorlinks=true, linkcolor=blue, citecolor=black, urlcolor=blue,
pdfborder={0 0 0}, pdfpagelabels]{hyperref}
\def\vx{\mathbf x}
\def\vy{\mathbf y}

\def\v0{\boldsymbol{0}}

\newlength{\FigureHeight}
\newlength{\FigureHeightHalf}

\numberwithin{equation}{section}
\titleformat{\section}
{\large\bfseries}{\thetitle.}{0.5em}{}
\titleformat{\subsection}
{\normalfont\bfseries}{\thetitle.}{0.5em}{}
\titleformat{\subsubsection}
{\normalfont\itshape}{\normalfont\thetitle.}{0.5em}{}
\newcommand{\footnoteref}[1]{\textsuperscript{\ref{#1}}}
\begin{document}

\title{\vspace{0.0em} Comment on `Conformal invariance of the zero-vorticity Lagrangian path in 2D turbulence'}
\author{Michael Frewer$\,^1$\thanks{Email address for correspondence:
frewer.science@gmail.com}$\:\,$ \& George Khujadze$\,^2$ \\ \\
\small $^1$ Heidelberg, Germany\\
\small $^2$ Chair of Fluid Mechanics, Universit\"at Siegen, 57068
Siegen, Germany}
\date{{\small\today}}
\clearpage \maketitle \thispagestyle{empty}

\vspace{-2.0em}\begin{abstract}

\noindent The current claim by Grebenev~{\it et~al.}~[\href{https://doi.org/10.1088/1751-8121/ab2f61}{J.~Phys.~A:~Math.~Theor.$\,$52,~335501~(2019)}], namely that the inviscid and unclosed 2D Lundgren-Monin-Novikov (LMN) equations on a zero-vorticity Lagrangian path admit conformal invariance, is based on a flawed and misleading analysis published earlier by \cite{Grebenev17}. All false results and conclusions made before in the Eulerian picture were now extended by \cite{Grebenev19} to the Lagrangian picture. Although we have already commented on these errors and consistently refuted their previous study \citep{Frewer18}, we deem it necessary to address and discuss these errors again in the new formulation and notation of \cite{Grebenev19} as it will offer new insights into this issue.

\vspace{0.5em}\noindent{\footnotesize{\bf Keywords:} {\it Statistical Physics, Conformal Invariance, Turbulence, Probability Density Functions, Lie Groups, Symmetry Analysis, Integro-Differential Equations,
Closure Problem}}
\end{abstract}

\section{Introduction and a remark on the notation\label{Sec1}}

The current publication by \cite{Grebenev19} is seriously flawed in the very same way as their previous one \citep{Grebenev17}. Their proposed analytical proof, namely that the (unclosed and non-modelled) PDF vorticity equations in the 2D inviscid turbulent flow case admit conformal invariance on a zero-vorticity characteristic, is false and misleading.

Despite the fact that the considered system of equations (2)-(4) in \cite{Grebenev19} for the considered case $n=1$ (see Sec.$\,$3) is unclosed and inherently would therefore allow for an unclosed set of invariances by itself, this system does not admit conformal invariance, neither in the Eulerian nor in the Lagrangian formulation, and this irrespective of whether a zero-vorticity characteristic is considered or not.

Note that when in the following all equation, section and page numbers in the present text appear as black, they refer to \cite{Grebenev19}, while all in blue refer to this comment, which here, of course, will be linked accordingly. Further note that when comparing the results between \cite{Grebenev17} and \cite{Grebenev19}, each is based on a different notation. The variables and functions
\begin{equation}
x^1,\; x^2,\; x^{\prime 1},\; x^{\prime 2},\; \xi^0,\; \xi^1,\; \xi^2,\; \xi^3,\; \xi^4,\; \xi^5,\; \xi^6,\; \eta^0=\eta_{f_1},\; \eta^\prime=\eta_{f_2},\; b^0,\; b^\prime,
\end{equation}
used in \cite{Grebenev17}, and likewise in our comment \cite{Frewer18}, were renamed in \cite{Grebenev19} to
\begin{equation}
x,\; y,\; x^\prime,\; y^\prime,\; \xi^t,\; \xi^x,\; \xi^y,\; \xi^\omega,\; \xi^{x^\prime},\; \xi^{y^\prime},\; \xi^{\omega^\prime},\; \eta^1,\; \eta^2,\; b_1,\; b_2,
\end{equation}
respectively. Also note here that $\eta^1$, $\eta^2$, $b_1$, $b_2$ in \cite{Grebenev19} do not correspond to $\eta^1$,~$\eta^2$,~$b^1$,~$b^2$ in \cite{Grebenev17}. They are different functions: While the former ones directly refer to the symmetry solutions of $f_1$ and $f_2$, the latter ones only refer to the symmetry solutions of the auxiliary (non-local) functions $J^1$ and $J^2$ as defined in \cite{Grebenev17},\linebreak[4] which then in turn defines the invariant transformations of the local functions $f_1$ and $f_2$.\linebreak[4] Furthermore, the composite variable $\vy=(\vx,\omega,\vx^\prime,\omega^\prime)$ as defined in \cite{Grebenev17} is not used anymore in \cite{Grebenev19}.

\section{Revealing the error using the smoothness-axiom of Lie-groups\label{Sec2}}

The heart of their non-correctable error in \cite{Grebenev19} lies in the interplay between result (30) and (24):
\begin{align*}
&(30)\!:\quad\; \xi^\omega=\big[6c^{11}(\vx)\big]\omega,\\[0.5em]
&(24)\!:\quad\; \frac{\partial\xi^\omega}{\partial x}=\frac{\partial\xi^\omega}{\partial y}=0.
\end{align*}
The solution constraint (24) says that the infinitesimal~$\xi^\omega$, given by (30), should not~depend on the spatial coordinates $x$ and $y$. Sure, at first glance (30) and (24) stand in conflict with each other, because to force a persistent spatial dependence with (30) is obviously not compatible with the spatial independence
as demanded by (24). But at a second glance, when particularly looking at the functional structure of (30), it seems that this conflict can be easily resolved if $\omega=0$ is chosen,
as the authors then did after equation (40).

Because now, with this specification $\omega=0$, result (30) turns to $\xi^\omega=0$ with which (24) then turns into $0=0$ and which therefore, according to the rationale of \cite{Grebenev19}, can be successfully removed from the invariance group simply because this constraint (24) gets identically satisfied when evaluated at $\omega=0$. That constraint (24) is indeed removed from the invariance group for $\omega=0$ can be explicitly seen, e.g., in the (incorrect)\footnote{Two separate and independent proofs will be provided in the next sections, that (45) and the mentioned statement in Sec.$\,$4.1 in \cite{Grebenev19} are indeed both false and misleading.\label{fn1}} final result (45), or in the (false and misleading)\footnoteref{fn1} statement in Sec.$\,$4.1 on p.12: {\it ``The invariance of the normalisation conditions (4) under the action of the group $G$ (25-31), (41) and (42) is evident and was derived before in [1]"}, which explicitly states that only (25-31), (41) and (42) are part of the group $G$ and not (24) anymore.

It's clear what the implications are when incorrectly removing constraint (24) from the\linebreak[4] invariance group for $\omega=0$: The function $c^{11}$ in (30) need not to be reduced to a global constant since it need not to comply with (24) anymore (due to its ``non-restricting" form $0=0$), but can remain to be a general function on the spatial coordinates $x$ and $y$, which then, along with the conditions (37-39), allows for the desired conformal invariance (26-29) not to get broken. But this reasoning is flawed and thus invalid as we will prove next.

\subsection{Proof that (24) may not be removed from the invariance group even if $\boldsymbol{\omega=0}$\label{Sec2.1}}

Even when putting $\omega=0$ in order to enforce compatibility between (24) and (30), and the constraint (24) itself pretends to be in the non-restrictive form $0=0$, the authors do not have a magic wand to simply let (24) disappear from the invariance group. The constraint (24) is still there and active even after putting $\omega=0$, simply because (24) is permanently valid for\linebreak[4] {\it all} real numbers of $\omega$, and that without any restrictions, which eventually is a crucial information not explicitly mentioned by the authors. But what does this additional information imply now? Well, since~(24) is {\it continuously} valid for {\it all} $\omega\in\mathbb{R}$ without any restrictions (which we will discuss at length further below and prove in detail in Appendix \ref{SecA}), we can take, for example, any differential consequence of (24) according to this variable without restrictions, for which we will then get further combined constraint equations also {\it continuously} valid for {\it all} $\omega\in\mathbb{R}$, on account of the underlying smoothness-axiom of Lie-groups. For example, due to the global existence of (24)
\begin{equation}
\frac{\partial\xi^\omega}{\partial x}=\frac{\partial\xi^\omega}{\partial y}=0, \;\;\forall \omega\in\mathbb{R},
\label{190713:1927}
\end{equation}
we thus~can imply the following differential consequence:
\begin{equation}
\frac{\partial^2 \xi^\omega}{\partial\omega\partial x}=\frac{\partial^2 \xi^\omega}{\partial\omega\partial y}=0, \;\;\forall \omega\in\mathbb{R}
\quad\;\;\Longleftrightarrow \quad\;\;
\frac{\partial}{\partial x}\left(\frac{\partial \xi^\omega}{\partial \omega}\right)=\frac{\partial}{\partial y}\left(\frac{\partial \xi^\omega}{\partial \omega}\right)=0, \;\;\forall \omega\in\mathbb{R}.
\label{190712:1446}
\end{equation}
The right-hand side of this implication \eqref{190712:1446} tells us now that the function $\xi_\omega^\omega:=\partial_\omega \xi^\omega$ should not depend on the spatial coordinates $x$ and $y$ for {\it any} value of $\omega$ as well, i.e.,
\begin{equation}
\frac{\partial\xi_\omega^\omega}{\partial x}=\frac{\partial\xi^\omega_\omega}{\partial y}=0, \;\;\forall \omega\in\mathbb{R},
\label{190712:1626}
\end{equation}
where initially in \eqref{190712:1446}, and this is important, we explicitly made use of the smoothness-axiom of Lie groups which allows for the interchanging of partial derivatives on its elements.

Hence, besides $\xi^\omega$, also $\xi^\omega_\omega$ should be spatially independent for  {\it any} value of $\omega$, including $\omega=0$.\linebreak[4] But this result causes a problem now. While the space-dependent result (30)
\begin{equation}
\xi^\omega=\big[6c^{11}(\vx)\big]\omega,
\label{190713:1935}
\end{equation}
could still be made compatible with constraint \eqref{190713:1927} by choosing $\omega=0$, this clearly does not work anymore for the next higher-order constraint \eqref{190712:1626}, since $\xi^\omega$ is linear in $\omega$.
Inserting \eqref{190713:1935} into \eqref{190712:1626} then leads to
\begin{equation}
\frac{\partial c^{11}(\vx)}{\partial x}=\frac{\partial c^{11}(\vx)}{\partial y}=0, \;\;\forall \omega\in\mathbb{R},
\label{190713:2318}
\end{equation}
which will reduce the spatial function $c^{11}$ to a global constant and thus, as a final result, the desired conformal invariance is broken, in particular also for $\omega=0$.\qed

\vspace{1em}
It's clear that the crucial aspect to obtain the above result \eqref{190712:1626} is that (24) has to be valid for {\it all} $\omega\in\mathbb{R}$, as explicitly and transparently written in \eqref{190713:1927}. Indeed, this unrestricted condition on constraint (24), namely that $\xi^\omega$ is globally independent on the spatial coordinates for any $\omega$,\linebreak[4] has already been consistently proven in our comment \cite{Frewer18} --- see~the result~(1.9) therein, where we even could prove that $\xi^\omega$ shows no other dependence than solely on $\omega$, and that this dependence is unrestricted, i.e., it is not constrained by any hidden or explicit condition on $\omega$:
\begin{equation}
\xi^\omega\equiv \xi^\omega(\omega),\: \text{unrestrictedly for~{\it all}~$\omega\in\mathbb{R}$}.
\end{equation}
In other words, when performing a thorough symmetry investigation, as we did in \cite{Frewer18}, it shows that the constraint (24) as given in \cite{Grebenev19} is not complete. It correctly has to be extended to:
\begin{equation}
\frac{\partial\xi^\omega}{\partial x}=\frac{\partial\xi^\omega}{\partial y}=\frac{\partial\xi^\omega}{\partial x^\prime}=\frac{\partial\xi^\omega}{\partial y^\prime}
=\frac{\partial\xi^\omega}{\partial \omega^\prime}=\frac{\partial\xi^\omega}{\partial t}=0, \;\;\forall \omega\in\mathbb{R}, \;\;\text{while}\;\; \frac{\partial\xi^\omega}{\partial \omega}\neq 0.
\label{190712:1711}
\end{equation}
To avoid any misunderstandings of this particular result and all previously obtained conclusions, the following should be~noted:

First, the constraint \eqref{190712:1711} is {\it not} an assumption, but the result of a thorough symmetry investigation, which we carefully checked both by hand as well as by using third-party computer software\footnote{In Appendix \ref{SecA} we provide an explicit proof of \eqref{190712:1711}, in that we perform a complete invariance analysis of the defining local equation that results to \eqref{190712:1711}.} that systematically calculates all Lie group symmetries of differential equations automatically. Because \eqref{190712:1711} is essentially nothing else than the full symmetry solution for the

\newgeometry{left=2.50cm,right=2.50cm,top=2.50cm,bottom=2.25cm,headsep=1em}

\noindent unknown $\xi^\omega$ of the local (differential) part of the considered system, which in \cite{Grebenev17} is given by equation (6). This is also indicated in \cite{Grebenev19} just before presenting their solution~(24). But they failed to give here the full symmetry solution \eqref{190712:1711}, which already was the case in their previous publication \citep{Grebenev17}.

Second, although the solution for the infinitesimal $\xi^\omega$ (30) can be viewed as a correct intermediate result, being a 3D-function of three independent variables, $\xi^\omega:=F(x,y,\omega)$, it\linebreak[4] nevertheless~unrestrictedly reduces to a 1D-function once it's subjected to its accompanying solution constraint $\partial_x\xi^\omega=\partial_y\xi^\omega=0$, $\forall \omega\in\mathbb{R}$ \eqref{190713:1927}, thus yielding: $\xi^\omega\equiv\xi^\omega(\omega)$, valid  $\forall\omega\in\mathbb{R}$,\linebreak[4] including the case $\omega=0$ --- a result which is ultimately rooted in the smoothness-axiom of Lie-groups as was shown~above.

Third, to already counteract in advance any opinions that incorrectly might say,
\vspace{-0.20em}
\begin{itemize}
\item[]\noindent {\it ``The logic of Frewer \& Khujadze is faulty, mainly because the symmetry operator (23) in Grebenev et al. (2019) seems to imply that the relevant derivatives one should consider
with respect to $\omega$, in order to establish differential consequences, are $\xi^\omega\partial_\omega$, and not just~$\partial_\omega$"},
\end{itemize}
\vspace{-0.25em}
here is our counterstatement, that clearly refutes any such opinion:

Initially (23) is an unknown symmetry operator to be determined such that it leaves invariant the considered system (2)-(6) for $n=1$. It defines a set of equations for the unknown infinitesimals, like the one for $\xi^\omega$, while (24)-(36) is a solution to these equations defined by~(23). So the intermediate result (30) is a solved and thus given solution function $F$ of three independent variables $\xi^\omega:=F(x,y,\omega)$, which now can be analyzed as one prefers. Hence, operator~(23) definitely gives no restriction as how $F$ can or should be analyzed. Again, $F$ is a solved and given 3D-function which, with the tools of calculus, can be examined or analyzed in any thinkable way.
One possible choice is to study the functional behaviour of $F$ when applying the operator $\partial_\omega$. Another choice would be to take any other derivative operator with respect to $\omega$, e.g., the one inspired
from (23), either $\xi^\omega\partial_\omega$, or the more general combined derivative operator $\xi^x\partial_x+\xi^y\partial_y+\xi^\omega\partial_\omega$, or, an operator which is completely different
to the ones just mentioned, e.g., the integral with respect to $\omega$, and so on, ad infinitum of possible choices of how one can analyze the behaviour of $F$. Of course, in each case one analyzes a different property of $F$.

Now, the aim in our first proof \eqref{190713:1927}-\eqref{190713:2318} was to study the property of $F$ if one just applies the operator $\partial_\omega$ to it. In this first proof we were not interested in what happens if we apply any other operator on $F$. Now, when merging (30) with the in parallel existing intermediate solution~(24), which unrestrictedly holds for all $\omega\in\mathbb{R}$ (see Appendix \ref{SecA} for a detailed proof), the correct analysis of $F$ with respect to $\partial_\omega$ leads straight to \eqref{190713:2318}, telling us that the conformal group is broken. Sure, when taking any other derivative operator than $\partial_\omega$, we will get different information about $F$, but this information is not relevant to us if it does not lead to \eqref{190713:2318}.\linebreak[4] For example, in our second proof in Sec.$\,$\ref{Sec3}, which is independent of the first proof \eqref{190713:1927}-\eqref{190713:2318}, we use the integral operator with respect to $\omega$, which is interesting again, because it also leads straight to \eqref{190713:2318} again. In this regard, please note that besides these two proofs shown in this comment, we still offer two more alternative proofs in \cite{Frewer18}, which all are independent of each other, and all leading to the same result, namely that there is no conformal group as claimed by Grebenev et al.

\subsection{Corollary: Result (45) as the final result for the invariance group is false\label{Sec2.2}}

Clearly result (45) is in error because it is not adhering to the higher-order constraint \eqref{190712:1626} of the invariance group for the considered system (2)-(3) for $n=1$. Only when including this constraint the correct result can be obtained --- higher order constraints beyond \eqref{190712:1626} are not needed since~the critical infinitesimal $\xi^\omega$ is at most linear in $\omega$. It is not surprising that the correct final invariance operator $\mathcal{S}$ in \cite{Grebenev19} should be given by
\begin{align}
\mathcal{S}&=\xi^x\partial_x+\xi^y\partial_y+2\xi^x_x\omega\partial_\omega-2\xi^x_xf_1\partial_{f_1}
+\xi^{x^\prime}\partial_{x^\prime}+\xi^{y^\prime}\partial_{y^\prime}
+\textstyle{\frac{2}{3}}\xi^x_x\omega^\prime\partial_{\omega^\prime}-\textstyle{\frac{8}{3}}\xi^x_xf_2\partial_{f_2},\nonumber\\[0.5em]
&\hspace{0.5cm}\text{with $\xi^x_x=3c^{11}$, and $\partial_x c^{11}=\partial_y c^{11}=0$},
\end{align}

\restoregeometry

\noindent that is, having the same operational structure as (45) but going along with the decisive extension that $c^{11}$ has to be a global constant not depending on the spatial coordinates $x$ and $y$, breaking thus the conformal symmetry of (45). The reason for why the operational structure of (45) stays unchanged when including the constraint \eqref{190712:1626} is that its effect on operator $\mathcal{S}$ is only to change the dependency of the function $c^{11}$ through \eqref{190713:2318}.

Furthermore, as this wrong invariance (45) has then been further used in \cite{Grebenev19} to demonstrate the conformal invariance for the evolution equations of the characteristic curves (11) and (12), it is clear that this demonstration, which was done explicitly in Sec.$\,$4 in the compact complex variable frame, is not valid and therefore misleading simply because this invariance does not exist for this system. It is not (11-12) itself which breaks the conformal invariance, but the evolution equation (2) of the PDFs $f_n$, necessary to evaluate the terms (14-15)\linebreak[4] for the evolution equations of the characteristic curves (11-12), which breaks~it.
The crucial constraints \eqref{190713:1927} and \eqref{190712:1626} arise because of the existence of (2), and not because of (11-12).

\section{The breaking of conformal invariance by the normalization condition\label{Sec3}}

In this section we present a second independent proof that refutes the claim of conformal invariance in \cite{Grebenev19}. In particular, their statement in Sec.$\,$4.1 on p.$\,$12 that {\it ``the~invariance of the normalisation conditions (4) under the action of the group $G$ (25-31), (41) and (42) is evident and was derived before in [1]"} is simply wrong, even in the rationale of \cite{Grebenev19}. In the following we will {\it not} use result~\eqref{190712:1626} to start this proof, as it would be too simple, but instead will proceed only with the information and reasoning as it is provided and presented in \cite{Grebenev19}, to demonstrate that if the authors would have done a thorough transformation of the normalization condition (4) along the lines of their own reasoning, they would have immediately realized that the conformal invariance need to get broken in order to have compatibility with (4). But such a thorough transformation on~(4) has not been done by them, and therefore they miss this crucial fact. Here we provide this analysis, forming parts of our complete survey given in \cite{Frewer18}.

In particular, we will now demonstrate that when transforming already the very first normalization condition in (4) for $n=1$, then it does {\it not} stay invariant under the group $G$ as proposed in \cite{Grebenev19}, which according to them consists only of the elements (25-31), (41) and (42). We start by asking that if this normalization condition is valid in the new transformed variables
\begin{equation}
\int d\omega^*f^*_1=1
\quad\;\;\Longleftrightarrow \quad\;\;
0 = 1-\int d\omega^*f^*_1,
\label{190714:1911}
\end{equation}
would it then stay invariant when transforming it back to its old variables? The $^*$-symbol above denotes the new variables which are connected to the old variables via the infinitesimal group transformation according to
(30) and (41) as given in \cite{Grebenev19}
\begin{equation}
\omega^*=\omega+\epsilon\cdot\xi^\omega +\mathcal{O}(\epsilon^2),\qquad f^*_1=f_1+\epsilon\cdot\eta^1 +\mathcal{O}(\epsilon^2),
\label{190714:1915}
\end{equation}
where $\epsilon\ll 1$ is the infinitesimal group parameter. Since \eqref{190714:1911} is a non-local relation in the variable $\omega$ to be transformed, we obviously need a transformation rule \eqref{190714:1915} for $\omega$ which is valid for {\it all} $\omega\in\mathbb{R}$, simply because \eqref{190714:1911} sums over {\it all} values of $\omega$ without any exceptions. Hence, we~need a $\xi^\omega$ which is valid for {\it all} $\omega\in\mathbb{R}$. Using only the information provided in \cite{Grebenev19}, as notably compiled on p.$\,$7, the infinitesimal $\xi^\omega$ can thus only be of the form
\begin{equation}
\xi^\omega=
\begin{cases}
\: 6c^{11}(\vx)\cdot\omega,\;\text{for $\omega=0$,}\\
\: \hspace{1.05cm}c\cdot\omega,\;\text{for $\omega\neq 0$, $c\neq 0$,}
\end{cases}
\label{190714:1937}
\end{equation}
where $c$ is some arbitrary constant. The function $c\cdot \omega$ for the case $\omega\neq0$ in \eqref{190714:1937} is a consequence of result (30)\footnote{Note that (30) is, as declared in \cite{Grebenev19}, the result when solving the non-local equations without any normalization. A detailed explanation is given in \cite{Grebenev17} --- see e.g. top of p.$\,$8.} to be compatible with the underlying constraint (24), such that $\xi^\omega$ is independent of the spatial coordinates $\vx=(x,y)$. Note that $\xi^\omega$ \eqref{190714:1937} is continuously differentiable at $\omega=0$,\linebreak[4] i.e., $\lim_{h\to 0} (\xi^\omega\vert_{\omega=0+h}-\xi^\omega\vert_{\omega=0})/h=\xi^\omega_\omega\big\vert_{\omega=0}$ exists, with $\lim_{\omega\to 0}\xi^\omega_\omega=\xi^\omega_\omega\vert_{\omega=0}$, and therefore the group element $\xi^\omega$ is not violating the smoothness-axiom of Lie-groups. Hence, for {\it all} $\omega\in\mathbb{R}$ the $\omega$-derivative of \eqref{190714:1937}~reads
\begin{equation}
\xi^\omega_\omega=c,\;\;\forall \omega\in\mathbb{R},
\label{190714:2029}
\end{equation}
which, of course, also includes the derivative at $\omega=0$, which again explicitly reads
\begin{align}
\xi^\omega_\omega\big\vert_{\omega=0}&=\lim_{h\to 0^-}\frac{\xi^\omega\vert_{\omega=0+h}-\xi^\omega\vert_{\omega=0}}{h}
=\lim_{h\to 0^+}\frac{\xi^\omega\vert_{\omega=0+h}-\xi^\omega\vert_{\omega=0}}{h}\nonumber\\[0.5em]
&=\,\lim_{h\to 0}\frac{c\cdot(0+h)-6c^{11}(\vx)\cdot 0}{h}\, =\, c.
\end{align}
In the rationale of \cite{Grebenev19}, the correct infinitesimal transformation rule to transform \eqref{190714:1911} is therefore given by \eqref{190714:1915}, where $\xi^\omega$ is given by \eqref{190714:1937} along with \eqref{190714:2029}, and $\eta^1$ by (41) in \cite{Grebenev19}: $\eta^1=-6c^{11}(\vx)f_1-(C_1+C_2)f_1+b_1(\vx,\omega,t)$. Now, let's transform \eqref{190714:1911} exactly according to this rule and see what happens:
\begin{align}
0 &= 1-\int d\omega^*f^*_1=1-\int d\omega\left|\frac{\partial\omega^*}{\partial\omega}\right|\big(f_1+\epsilon \eta^1+\mathcal{O}(\epsilon^2)\big)
=1-\int d\omega \big|1+\epsilon\xi^\omega_\omega\big|\big(f_1+\epsilon \eta^1\big)+\mathcal{O}(\epsilon^2)\nonumber\\[0.5em]
&\underset{\epsilon\ll 1}{=}1-\int d\omega \big(1+\epsilon\xi^\omega_\omega\big)\big(f_1+\epsilon \eta^1\big)+\mathcal{O}(\epsilon^2)
=1-\int d\omega \big(f_1+\epsilon\big(\eta^1+\xi^\omega_\omega f_1\big)\big) +\mathcal{O}(\epsilon^2)\nonumber\\[0.5em]
&=\underbrace{1-\int d\omega \big(f_1}_{=\,0}+\epsilon\big(-6c^{11}(\vx)f_1-(C_1+C_2)f_1+b_1(\vx,\omega,t)+cf_1\big)\big) +\mathcal{O}(\epsilon^2)\nonumber\\[0.0em]
&=\underbrace{\big(6c^{11}(\vx)+C_1+C_2-c\big)}_{\text{constant in $\omega$}}\underbrace{\int d\omega f_1}_{=\,1}
-\underbrace{\int d\omega b_1(\vx,\omega,t)}_{=\,C_1+C_2,\;\text{see (44)\footnotemark}} + \mathcal{O}(\epsilon) \nonumber\\[0.0em]
&=\big(6c^{11}(\vx)-c\big)+\mathcal{O}(\epsilon),
\label{190712:2317}
\end{align}
\footnotetext{The second integral result in (44) in \cite{Grebenev19} misses the constant $C_1$ next to $C_2$. The problem is that the authors distributed the constant $C_1$ in their new version differently than in their earlier version, where $C_1$ was also associated to the solution of $\xi^\omega$ (see result (38) in \cite{Grebenev17}), while in the new version~not. Nevertheless, all the different redistributions of these constants have no effect on the proof given above. In our comment \cite{Frewer18} we even perform this proof \eqref{190712:2317} on a more general basis, since a complete and thorough symmetry analysis shows that $C_1$ and $C_2$ may also depend on $\omega$, with the consequence then that they may not be pulled in front of the $\omega$-integrations anymore as it was done in \eqref{190712:2317}. But also with this generalized proof (see Sec.$\,$2.2 therein) we come of course to the same conclusion in that the proposed conformal symmetry is not compatible with the normalization condition.}
which can only be satisfied if
\begin{equation}
6c^{11}(\vx)-c=0,\;\;\forall \vx\in\mathbb{R}^2,
\label{200502:1819}
\end{equation}
that is, if and only if $c^{11}$ is a constant for all coordinates $\vx$, which of course breaks the conformal invariance.\footnote{Note that \eqref{200502:1819} could already have been obtained by taking the naive $\omega$-derivative of $\xi^\omega$~\eqref{190714:1937}, with result $\xi^\omega_\omega=6c^{11}(\vx)$, for $\omega=0$, and $\xi^\omega_\omega=c$, for $\omega\neq0$, and then by enforcing the axiom of continuity to $\xi^\omega_\omega$ at~$\omega=0$.} Hence, we obtain the correct final result that the normalization condition~(4) is only compatible to the considered symmetry group $G$ if it contains a spatially {\it independent} function~$c^{11}$, which of course is completely opposite as to what is claimed in Sec.$\,$4.1 on p.12 in \cite{Grebenev19} by wrongly allowing for a spatial dependence in $c^{11}$.\qed

\vspace{1em}
Note that it's not a minor issue that the normalization condition (4) breaks the conformal invariance, e.g., by saying then let's ignore the normalization condition from the system in order to restore this invariance. The normalization condition cannot be ignored, because it's an internal condition that guarantees that any PDF solution $f_n$ stays physically valid during evolution, or as \cite{Grebenev19} correctly puts it: {\it ``Physically meaningful fields satisfy the properties (4-6)"} [Sec.$\,$4.1, p.12]. In other words, if an invariance operator for the PDF system,


\noindent as given in \cite{Grebenev19} by (2), is not compatible to the normalization condition (4), then physical solutions can get mapped to unphysical ones. Therefore the normalization is an important ingredient in any PDF system and should be respected within a symmetry analysis.\linebreak[4] The same is true for all other internal constraints that go along with such a PDF system, all necessary to ensure physical PDF solutions. In this regard, please see our supplementing comment \cite{Frewer17}, which criticizes an even earlier publication by V.~Grebenev \citep{Waclawczyk17}, in that new invariance groups get proposed therein which obviously are not compatible to the full PDF system when including all internal constraints, i.e., ultimately, in \cite{Waclawczyk17} non-physical symmetries are getting proposed.

\section{Final remarks\label{Sec4}}

\noindent {\bf R1.} It should be clear that our comment did not question the (possible) existence of conformal invariance in 2D turbulence, as e.g. indicated by \cite{Bernard06}.
What is criticized and refuted herein is only the algebraic derivation by
Grebenev et al. and their simplistic idea that the conformal invariance group would naturally arise from the first-order unclosed PDF-formulation of the 2D (inviscid) Navier-Stokes equations when only analyzing these by means of a classical Lie-group symmetry approach. In fact, as we have proven, their algebraic derivation for conformal invariance of the 2D LMN vorticity equations is flawed in both the Eulerian \citep{Grebenev17} as well as in the Lagrangian picture \citep{Grebenev19}.

\noindent{\bf R2.} The following statement in \cite{Grebenev19} on p.$\,$8 that {\it ``these relationships [(40)] explicitly demonstrate the exceptional role of the zero-vorticity constraint
$\omega=0$ to guarantee that $c^{11}$ and $c^{22}$ are non-trivial functions and the CG [conformal group] appears for $\omega=0$"} is~not only false but also seriously misleading. In Sec.$\,$2.3 in \cite{Frewer18} we clearly demonstrate that the choice $\omega=0$ is not exceptional at all, because the obtained invariances can always be equivalently re-formulated such that {\it any} arbitrary but fixed value of $\omega$ will do the same job as the particular choice $\omega=0$ --- see therein particularly our result (2.28) for an alternative~$\xi^\omega$ and its subsequent discussion. Hence, opposite to their claim, the choice of a zero-vorticity constraint $\omega=0$ plays no exceptional role, resulting even in the fact that the proposed conformal invariance is not only broken for $\omega=0$, but for all $\omega\in\mathbb{R}$, thus refuting \cite{Grebenev17,Grebenev19} in its most general form.

\noindent {\bf R3.} Important to note in this overall discussion is that all invariant transformations put forward in \cite{Grebenev19} are only equivalence and not true symmetry transformations, simply due to that we are dealing here with an unclosed system of equations (2)-(3), where, for $n=1$, the dynamical rule of the 2-point PDF $f_2$ is not known beforehand.
In contrast to a true symmetry transformation, which maps a solution of a specific (closed) equation to a new solution of the same equation, an equivalence transform acts in a weaker sense in that it only maps an (unclosed) equation to a new (unclosed) equation of the same class.\footnote{Equivalence transformations can be successfully applied for example to classify unclosed differential equations according to the number of symmetries they admit when specifying the unclosed terms (see e.g. \cite{Meleshko02,Khabirov02.1,Khabirov02.2,Chirkunov12,Meleshko15,Bihlo17}). A typical task in this context sometimes is to find a specification of the unclosed terms such that the maximal symmetry algebra is gained. Once the equation is closed by a such a group classification, invariant solutions can be determined. But in how far these equations and their solutions are physically relevant and whether they can be matched to empirical data is not clarified {\it a priori} by this approach, in particular if such a pure Lie-group-based type of modelling is performed fully detached from empirical research. In this regard, special attention has to be given to the unclosed statistical equations of turbulence as considered herein, since the unclosed 2-point PDF in (2)-(6) in \cite{Grebenev19} for $n=1$, is only an analytical and theoretical unknown, but not an empirical one since it is fully determined by the underlying deterministic Navier-Stokes equations, which again are well-known for to break statistical symmetries in turbulence within intermittent events (see e.g. \cite{Frisch95}). Hence extra caution has to be exercised when employing a pure symmetry-based modeling to turbulence.}

\noindent Of course, it is trivial and goes without saying that if once a real solution for $f_2$ is known,
and if the equivalence (42) itself (for $c^{11}_1=c^{11}_2=0$) is physically realizable,\footnote{The equivalence transformation (42) (for $c^{11}_1=c^{11}_2=0$) is physically realizable only if the transformed field $f^*_2$ can be generated as a PDF-solution of the deterministic Navier-Stokes equations according to its transformation rule $f^*_2=f_2+\epsilon\cdot\eta^2+\mathcal{O}(\epsilon^2)$, where we assume that the
non-transformed field $f_2$ already constitutes a PDF-solution.
In other words, if $f^*_2$ cannot emerge dynamically from $f_2$ via the deterministic and thus closed Navier-Stokes equations, then the equivalence (42) (for $c^{11}_1=c^{11}_2=0$) is nonphysical.
To 
prove whether this equivalence is physically realizable or not, is beyond the scope of this article.
However, there are a few examples of statistical Navier-Stokes equivalences which are clearly nonphysical --- see e.g. \cite{Frewer14.2,Frewer15,Frewer16,Frewer17,Sadeghi20}.}
then this equivalence turns into a symmetry transformation and $f_2$ gets mapped to a new solution $f^*_2=f_2+\epsilon\cdot\eta^2+\mathcal{O}(\epsilon^2)$. But since this is not the case here, any valid invariant transformation in (24)-(45) (for $c^{11}_1=c^{11}_2=0$) will thus at this stage only map between equations and not between solutions, where $f_2$ then is the unknown source or sink term, or collectively the unknown constitutive law of these equations.

Hence, even for all invariant transformations that still remain valid in \cite{Grebenev19}, we cannot expect any information about the inner solution structure of the 1-point PDF equation as long as the dynamical equation for the 2-point PDF $f_2$ is not modeled. Without empirical modeling it is clear that the closure problem of turbulence cannot be circumvented by just employing the method of a Lie-group symmetry analysis. For more details on this issue, see e.g. \cite{Frewer14.1,Frewer14.2} and the references therein.


\appendix

\section{The general and full invariance group of the local equation\label{SecA}}

In the following we will explicitly prove that solution (24) in \cite{Grebenev19} is not complete, in that it particularly misses the decisive and crucial information $\forall\omega\in\mathbb{R}$. In other words, we will prove that solution (24) has to be extended to the more general and true solution~\eqref{190712:1711}.

As declared in \cite{Grebenev19} that (24) is the solution of the local part of the problem, which in their earlier study \citep{Grebenev17} is explicitly given by equation (6), and in \cite{Frewer18}
by $\mathsf{E}_1$ (1.4), our proof here is thus based on determining the most general invariant solution of the local equation $\mathsf{E}_1$ (1.4). We will present two different but equivalent versions as how one can perform a systematic Lie-group invariance analysis on  $\mathsf{E}_1$~(1.4) using a software package, the DESOLV-II package of \cite{Vu12}.

In the first version (Version~No.$\,$1), we consider the three dependent variables $J^0$, $J^1$, $J^2$ in  $\mathsf{E}_1$~(1.4) to explicitly depend on all independent variables of the system involved. These are seven in total and are listed in \cite{Frewer18} by (1.3). Although we only consider here the local equation  $\mathsf{E}_1$~(1.4) and ignore in this step the non-local equations $\mathsf{E}_2$-$\mathsf{E}_5$~(1.5)-(1.8), we nevertheless should provide the symmetry searching algorithm with the information that three independent variables, namely $y^4=x^\prime$, $y^5=y^\prime$, $y^6=\omega^\prime$, are integration variables. This can be done by augmenting the local equation~$\mathsf{E}_1$~(1.4) with first-order differential consequences consistent with all equations $\mathsf{E}_1$-$\mathsf{E}_5$~(1.4)-(1.8) defining the system. The relevant ones are given by the system of equations $\partial_{y^j}J^k=0$, for all $k=0,1,2$ and $j=4,5,6$, obviously telling us that all three dependent variables $J^0$, $J^1$, $J^2$ do not explicitly dependent on the three (integration) variables $y^4$, $y^5$, $y^6$, which brings us then to the second~version.

In the second version (Version~No.$\,$2), we consider the three dependent variables $J^0$, $J^1$, $J^2$\linebreak[4] in $\mathsf{E}_1~$(1.4) to explicitly depend only on those independent variables which the full system\linebreak[4]
$\mathsf{E}_1$-$\mathsf{E}_5$~(1.4)-(1.8) defines for them. As established in the first version above, they thus can only dependent on the four variables $y^0$, $y^1$, $y^2$, $y^3$. Hence, the symmetry analysis of this version will only involve a single equation, the local equation $\mathsf{E}_1$~(1.4) itself.

As the computer results show below, {\it both} versions obviously yield the same final result (1.9)-(1.11) in \cite{Frewer18}, in that the infinitesimal $\xi^\omega=\xi^3$ is only a function of $\omega$, and this without any restrictions on the values of $\omega$, as stated correctly in this manuscript by~\eqref{190712:1711}.

\subsection{ Version No.$\,$1}
\input{V1}

\subsection{ Version No.$\,$2}
\input{V2}
\bibliographystyle{jfm}
\bibliography{References}

\end{document}

%% file: V1.tex
\begin{maplegroup}
\begin{flushleft}
{\large Header:}
\end{flushleft}
\end{maplegroup}
\begin{maplegroup}
\begin{mapleinput}
\mapleinline{active}{1d}{restart: read "Desolv-V5R5.mpl": with(desolv):}{}
\end{mapleinput}
\mapleresult
\begin{maplelatex}
\mapleinline{inert}{2d}{`DESOLVII_V5R5 (March-2011)(c) by Dr. K.
T. Vu, Dr. J. Carminati and
Miss. G. Jefferson`;}{%
\maplemultiline{ \mathit{\phantom{xxxxxxxx} DESOLVII\_V5R5\ (March-2011)(c)} \\
\mathit{by\ Dr.\ K.\ T.\ Vu,\ Dr.\ J.\ Carminati\ and\ Miss.\ G.\
Jefferson} }}
\end{maplelatex}
\end{maplegroup}
\begin{maplegroup}
\begin{flushleft}
{\large Definitions of variables, local equation and differential consequences:}
\end{flushleft}
\end{maplegroup}
\begin{maplegroup}
\begin{mapleinput}
\mapleinline{active}{1d}{alias(sigma=(y0,y1,y2,y3,y4,y5,y6,J0,J1,J2)): Y:=(y0,y1,y2,y3,y4,y5,y6):
}{}
\end{mapleinput}
\end{maplegroup}
\begin{maplegroup}
\begin{mapleinput}
\mapleinline{active}{1d}{eqn0:=diff(J0(Y),y0)+diff(J1(Y),y1)+diff(J2(Y),y2)=0:
}{}
\end{mapleinput}
\end{maplegroup}
\begin{maplegroup}
\begin{mapleinput}
\mapleinline{active}{1d}{eqn1:=diff(J0(Y),y4)=0: eqn2:=diff(J0(Y),y5)=0: eqn3:=diff(J0(Y),y6)=0:
\hspace{0.43cm} eqn4:=diff(J1(Y),y4)=0: eqn5:=diff(J1(Y),y5)=0: eqn6:=diff(J1(Y),y6)=0:
\hspace{0.43cm} eqn7:=diff(J2(Y),y4)=0: eqn8:=diff(J2(Y),y5)=0: eqn9:=diff(J2(Y),y6)=0:
}{}
\end{mapleinput}
\end{maplegroup}
\begin{maplegroup}
\begin{mapleinput}
\mapleinline{active}{1d}{eqns:=[eqn0,eqn1,eqn2,eqn3,eqn4,eqn5,eqn6,eqn7,eqn8,eqn9]:
}{}
\end{mapleinput}
\end{maplegroup}
\begin{maplegroup}
\begin{flushleft}
{\large Symmetry Algorithm:}\\[0.75em]
\textit{{\large Size of the determining system:}}
\end{flushleft}
\end{maplegroup}
\begin{maplegroup}
\begin{mapleinput}
\mapleinline{active}{1d}{detsys:=gendef(eqns,[J0,J1,J2],[y0,y1,y2,y3,y4,y5,y6]): nops(detsys[1]);
}{}
\end{mapleinput}
\mapleresult
\begin{maplelatex}
\mapleinline{inert}{2d}{45}{\[\displaystyle 45\]}
\end{maplelatex}
\end{maplegroup}
\vspace{-0.75em}
\begin{maplegroup}
\begin{flushleft}
\textit{{\large Solving the determining system:}}
\end{flushleft}
\end{maplegroup}
\begin{maplegroup}
\begin{mapleinput}
\mapleinline{active}{1d}{sym:=pdesolv(op(detsys));
}{}
\end{mapleinput}
\mapleresult
\begin{maplelatex}
\mapleinline{inert}{2d}{}{\[\displaystyle
{\it sym}\, := \,\bigg[\bigg[-{\frac {\partial }{\partial {\it y1}}}{\it F\_15} \left( {\it y0},{\it y1},{\it y2},{\it y3}\\ \mbox{} \right)
-{\frac {\partial }{\partial {\it y2}}}{\it F\_21} \left( {\it y0},{\it y1},{\it y2},{\it y3}\\
\mbox{} \right) -{\it F\_47} \left( {\it y0},{\it y1},{\it y2},{\it y3}\\ \mbox{} \right),
\]}
\end{maplelatex}
\mapleresult
\begin{maplelatex}
\mapleinline{inert}{2d}{}{\[\displaystyle
{\frac {\partial }{\partial {\it y0}}}{\it F\_44} \left( {\it y0},{\it y1},{\it y2},{\it y3}\\
\mbox{} \right)
+{\frac {\partial }{\partial {\it y1}}}{\it F\_45} \left( {\it y0},{\it y1},{\it y2},{\it y3}\\
\mbox{} \right)
+{\frac {\partial }{\partial {\it y2}}}{\it F\_46} \left( {\it y0},{\it y1},{\it y2},{\it y3}\\
\mbox{} \right) \\
\mbox{} \bigg],\,[\,],
\]}
\end{maplelatex}
\mapleresult
\begin{maplelatex}
\mapleinline{inert}{2d}{}{\[\displaystyle
\bigg[\xi_{{{\it y0}}} \left( {\it \sigma}\right)\!=\!{\it F\_27}\\
\mbox{} \left( {\it y0},{\it y1},{\it y2},{\it y3} \right),
\xi_{{{\it y1}}} \left( {\it \sigma}\right)\!=\!{\it F\_15} \left( {\it y0},{\it y1},{\it y2},{\it y3} \right),
\xi_{{{\it y2}}} \left( {\it \sigma}\right)\\
\mbox{}\!=\!{\it F\_21} \left( {\it y0},{\it y1},{\it y2},{\it y3} \right),
\]}
\end{maplelatex}
\mapleresult
\begin{maplelatex}
\mapleinline{inert}{2d}{}{\[\displaystyle
\xi_{{{\it y3}}} \left( {\it \sigma}\right)\!=\!{\it F\_9} \left( {\it y3} \right),
\xi_{{{\it y4}\\
\mbox{}}} \left( {\it \sigma}\right) \\
\mbox{}\!=\!\xi_{{{\it y4}\\
\mbox{}}} \left( {\it \sigma}\right),
\xi_{{{\it y5}}} \left( {\it \sigma}\right)\!=\!\xi_{{{\it y5}}} \left( {\it \sigma}\right),
\xi_{{{\it y6}}} \left( {\it \sigma}\right)\!=\!\xi_{{{\it y6}}} \left( {\it \sigma}\right),
\]}
\end{maplelatex}
\mapleresult
\begin{maplelatex}
\mapleinline{inert}{2d}{}{\[\displaystyle
\eta_{{{\it J0}}} \left( {\it \sigma}\right)\!=\!
{\it F\_47} \left( {\it y0},{\it y1},{\it y2},{\it y3} \right){\it J0}
\!+\!{\it J1}\,{\frac {\partial }{\partial {\it y1}}}{\it F\_27}\\
\mbox{} \left( {\it y0},{\it y1},{\it y2},{\it y3} \right)
\]}
\end{maplelatex}
\mapleresult
\vspace{-0.5em}
\begin{maplelatex}
\mapleinline{inert}{2d}{}{\[\displaystyle
\hspace{1.5cm}+{\it J2}\,{\frac {\partial }{\partial {\it y2}}}{\it F\_27}\\
\mbox{} \left( {\it y0},{\it y1},{\it y2},{\it y3} \right)\!+\!{\it F\_44} \left( {\it y0},{\it y1},{\it y2},{\it y3} \right) \\
\mbox{},
\]}
\end{maplelatex}
\vspace{-0.5em}
\mapleresult
\begin{maplelatex}
\mapleinline{inert}{2d}{}{\[\displaystyle
\eta_{{{\it J1}}} \left( {\it \sigma}\right)\!=\!
{\it J0}\,{\frac {\partial }{\partial {\it y0}}}{\it F\_15} \left( {\it y0},{\it y1},{\it y2},{\it y3} \right) \\
\mbox{}\!+\!{\it F\_47} \left( {\it y0},{\it y1},{\it y2},{\it y3} \right) {\it J1}
\!+\!{\it J2}\,{\frac {\partial }{\partial {\it y2}}}{\it F\_15} \left( {\it y0},{\it y1},{\it y2},{\it y3} \right)
\]}
\end{maplelatex}
\mapleresult
\vspace{-0.25em}
\begin{maplelatex}
\mapleinline{inert}{2d}{}{\[\displaystyle
\hspace{1.5cm}+{\it J1}\,{\frac {\partial }{\partial {\it y1}}}{\it F\_15} \left( {\it y0},{\it y1},{\it y2},{\it y3} \right)
\!-\!{\it J1}\,{\frac {\partial }{\partial {\it y0}}}{\it F\_27}\\
\mbox{} \left( {\it y0},{\it y1},{\it y2},{\it y3} \right)\!+\!{\it F\_45} \left( {\it y0},{\it y1},{\it y2},{\it y3} \right) ,
\]}
\end{maplelatex}
\mapleresult
\begin{maplelatex}
\mapleinline{inert}{2d}{}{\[\displaystyle
\eta_{{{\it J2}}} \left( {\it \sigma}\right) \\
\mbox{}\!=\!
{\it J0}\,{\frac {\partial }{\partial {\it y0}}}{\it F\_21} \left( {\it y0},{\it y1},{\it y2},{\it y3} \right)
\!+\!\,{\it F\_47} \left( {\it y0},{\it y1},{\it y2},{\it y3} \right){\it J2}
\!+\!{\it J2}\,{\frac {\partial }{\partial {\it y2}}}{\it F\_21} \left( {\it y0},{\it y1},{\it y2},{\it y3} \right)
\]}
\end{maplelatex}
\mapleresult
\begin{maplelatex}
\mapleinline{inert}{2d}{}{\[\displaystyle
\hspace{1.5cm}+{\it J1}\,{\frac {\partial }{\partial {\it y1}}}{\it F\_21} \left( {\it y0},{\it y1},{\it y2},{\it y3} \right)
\!-\!{\it J2}\,{\frac {\partial }{\partial {\it y0}}}{\it F\_27}\\
\mbox{} \left( {\it y0},{\it y1},{\it y2},{\it y3} \right)
\!+\!{\it F\_46} \left( {\it y0},{\it y1},{\it y2},{\it y3} \right)\\
\mbox{}\bigg],
\]}
\end{maplelatex}
\mapleresult
\begin{maplelatex}
\mapleinline{inert}{2d}{}{\[\displaystyle
\bigg[{\it F\_15} \left( {\it y0},{\it y1},{\it y2},{\it y3}\\
\mbox{} \right) ,{\it F\_21} \left( {\it y0},{\it y1},{\it y2},{\it y3}\\
\mbox{} \right) ,{\it F\_27} \left( {\it y0},{\it y1},{\it y2},{\it y3}\\
\mbox{} \right) ,{\it F\_44} \left( {\it y0},{\it y1},{\it y2},{\it y3}\\
\mbox{} \right), \\
\mbox{}
\]}
\end{maplelatex}
\mapleresult
\begin{maplelatex}
\mapleinline{inert}{2d}{}{\[\displaystyle
{\it F\_45} \left( {\it y0},{\it y1},{\it y2},{\it y3}\\
\mbox{} \right) ,{\it F\_46} \left( {\it y0},{\it y1},{\it y2},{\it y3}\\
\mbox{} \right) ,{\it F\_47} \left( {\it y0},{\it y1},{\it y2},{\it y3}\\
\mbox{} \right) \\
\mbox{},
{\it F\_9} \left( {\it y3}\\
\mbox{} \right),
\]}
\end{maplelatex}
\mapleresult
\begin{maplelatex}
\mapleinline{inert}{2d}{}{\[\displaystyle
\xi_{{{\it y4}}} \left( {\it \sigma}\right),
\xi_{{{\it y5}}} \left( {\it \sigma}\right),
\xi_{{{\it y6}\\
\mbox{}}} \left( {\it \sigma}\right) \\
\mbox{}\bigg]\bigg]
\]}
\end{maplelatex}
\end{maplegroup}
\begin{maplegroup}
\begin{flushleft}
{\large Redefining solution functions as used in (1.9)-(1.11):}
\end{flushleft}
\end{maplegroup}
\begin{maplegroup}
\begin{mapleinput}
\mapleinline{active}{1d}{F_27(y0,y1,y2,y3):=xi0(y0,y1,y2,y3); F_15(y0,y1,y2,y3):=xi1(y0,y1,y2,y3);
\hspace{0.43cm} F_21(y0,y1,y2,y3):=xi2(y0,y1,y2,y3); F_9(y3):=xi3(y3);
\hspace{0.43cm} F_47(y0,y1,y2,y3):=-diff(F_15(y0,y1,y2,y3),y1)-diff(F_21(y0,y1,y2,y3),y2);
\hspace{0.43cm} F_44(y0,y1,y2,y3):=b0(y0,y1,y2,y3)-C(y3)*j0(y0,y1,y2,y3);
\hspace{0.43cm} F_45(y0,y1,y2,y3):=b1(y0,y1,y2,y3)-C(y3)*j1(y0,y1,y2,y3);
\hspace{0.43cm} F_46(y0,y1,y2,y3):=b2(y0,y1,y2,y3)-C(y3)*j2(y0,y1,y2,y3);
}{}
\end{mapleinput}
\mapleresult
\begin{maplelatex}
\mapleinline{inert}{2d}{}{\[\displaystyle
{\it F\_27} \left({\it y0} ,{\it y1} ,{\it y2} ,{\it y3} \right):={\it \xi 0}\\
\mbox{} \left({\it y0} ,{\it y1} ,{\it y2} ,{\it y3} \right)
\]}
\end{maplelatex}
\mapleresult
\begin{maplelatex}
\mapleinline{inert}{2d}{}{\[\displaystyle
{\it F\_15} \left({\it y0} ,{\it y1} ,{\it y2} ,{\it y3} \right):={\it \xi 1}\\
\mbox{} \left({\it y0} ,{\it y1} ,{\it y2} ,{\it y3} \right)
\]}
\end{maplelatex}
\mapleresult
\begin{maplelatex}
\mapleinline{inert}{2d}{}{\[\displaystyle
{\it F\_21} \left({\it y0} ,{\it y1} ,{\it y2} ,{\it y3} \right):={\it \xi 2}\\
\mbox{} \left({\it y0} ,{\it y1} ,{\it y2} ,{\it y3} \right)
\]}
\end{maplelatex}
\mapleresult
\begin{maplelatex}
\mapleinline{inert}{2d}{}{\[\displaystyle
\hspace{0.20cm}{\it F\_9} \left({\it y3} \right):={\it \xi 3}\\
\mbox{} \left({\it y3} \right)
\]}
\end{maplelatex}
\mapleresult
\begin{maplelatex}
\mapleinline{inert}{2d}{}{\[\displaystyle
{\it F\_47} \left({\it y0} ,{\it y1} ,{\it y2} ,{\it y3} \right):=-\frac{\partial }{\partial {\it y1} }~{\it \xi 1}\\
\mbox{} \left({\it y0} ,{\it y1} ,{\it y2} ,{\it y3} \right)-\frac{\partial }{\partial {\it y2} }~{\it \xi 2}\\
\mbox{} \left({\it y0} ,{\it y1} ,{\it y2} ,{\it y3} \right)
\]}
\end{maplelatex}
\mapleresult
\begin{maplelatex}
\mapleinline{inert}{2d}{}{\[\displaystyle
{\it F\_44} \left({\it y0} ,{\it y1} ,{\it y2} ,{\it y3} \right):={\it b0}\\
\mbox{} \left({\it y0} ,{\it y1} ,{\it y2} ,{\it y3} \right)-C \left({\it y3} \right)~{\it j0} \left({\it y0} ,{\it y1} ,{\it y2} ,{\it y3} \right)
\]}
\end{maplelatex}
\mapleresult
\begin{maplelatex}
\mapleinline{inert}{2d}{}{\[\displaystyle
{\it F\_45} \left({\it y0} ,{\it y1} ,{\it y2} ,{\it y3} \right):={\it b1}\\
\mbox{} \left({\it y0} ,{\it y1} ,{\it y2} ,{\it y3} \right)-C \left({\it y3} \right)~{\it j1} \left({\it y0} ,{\it y1} ,{\it y2} ,{\it y3} \right)
\]}
\end{maplelatex}
\mapleresult
\begin{maplelatex}
\mapleinline{inert}{2d}{}{\[\displaystyle
{\it F\_46} \left({\it y0} ,{\it y1} ,{\it y2} ,{\it y3} \right):={\it b2}\\
\mbox{} \left({\it y0} ,{\it y1} ,{\it y2} ,{\it y3} \right)-C \left({\it y3} \right)~{\it j2} \left({\it y0} ,{\it y1} ,{\it y2} ,{\it y3} \right)
\]}
\end{maplelatex}
\end{maplegroup}
\begin{maplegroup}
\begin{flushleft}
{\large The arbitrary integration functions $b_k$ and $j_k$ are solutions of the local equation (1.4).\\
Since (1.4) is a linear equation: if $b_k$ and $j_k$ are solutions, so is $B_k:=b_k-C\cdot j_k$.}
\end{flushleft}
\end{maplegroup}
\begin{maplegroup}
\begin{mapleinput}
\mapleinline{active}{1d}{simplify(sym[1,1]); simplify(sym[1,2])=0;}{}
\end{mapleinput}
\mapleresult
\begin{maplelatex}
\mapleinline{inert}{2d}{0}{\[\displaystyle 0\]}
\end{maplelatex}
\mapleresult
\begin{maplelatex}
\mapleinline{inert}{2d}{}{\[\displaystyle
{\frac {\partial }{\partial {\it y0}}}{\it b0} \left( {\it y0},{\it y1},{\it y2},{\it y3} \right)
+{\frac {\partial }{\partial {\it y1}}}{\it b1} \left( {\it y0},{\it y1},{\it y2},{\it y3} \right)
+{\frac {\partial }{\partial {\it y2}}}{\it b2} \left( {\it y0},{\it y1},{\it y2},{\it y3} \right)
\]}
\end{maplelatex}
\mapleresult
\begin{maplelatex}
\mapleinline{inert}{2d}{}{\[\displaystyle
-C \left( {\it y3} \right)\left(
{\frac {\partial }{\partial {\it y0}}}{\it j0} \left( {\it y0},{\it y1},{\it y2},{\it y3} \right)
+{\frac {\partial }{\partial {\it y1}}}{\it j1} \left( {\it y0},{\it y1},{\it y2},{\it y3} \right)
+{\frac {\partial }{\partial {\it y2}}}{\it j2} \left( {\it y0},{\it y1},{\it y2},{\it y3} \right)
\right)=0
\]}
\end{maplelatex}
\end{maplegroup}
%
%
\begin{maplegroup}
\begin{flushleft}
{\large The solutions $j^k$ can be identified as the dependent variables $J^k$:}
\end{flushleft}
\end{maplegroup}
\begin{maplegroup}
\begin{mapleinput}
\mapleinline{active}{1d}{j0(y0,y1,y2,y3):=J0: j1(y0,y1,y2,y3):=J1: j2(y0,y1,y2,y3):=J2:
}{}
\end{mapleinput}
\end{maplegroup}
\begin{maplegroup}
\begin{flushleft}
{\large Final result which is identical to (1.9)-(1.11):}
\end{flushleft}
\end{maplegroup}
\begin{maplegroup}
\begin{mapleinput}
\mapleinline{active}{1d}{simplify(sym[3,1]); simplify(sym[3,2]); simplify(sym[3,3]);
\hspace{0.43cm} simplify(sym[3,4]); simplify(sym[3,8]); simplify(sym[3,9]);
\hspace{0.43cm} simplify(sym[3,10]);
}{}
\end{mapleinput}
\mapleresult
\begin{maplelatex}
\mapleinline{inert}{2d}{}{\[\displaystyle \xi_{{{\it y0}}} \left( {\it \sigma}\right)
={\it \xi 0}\\
\mbox{} \left( {\it y0},{\it y1},{\it y2},{\it y3} \right) \]}
\end{maplelatex}
\mapleresult
\begin{maplelatex}
\mapleinline{inert}{2d}{}{\[\displaystyle \xi_{{{\it y1}}} \left( {\it \sigma}\right) ={\it \xi 1}\\
\mbox{} \left( {\it y0},{\it y1},{\it y2},{\it y3} \right) \]}
\end{maplelatex}
\mapleresult
\begin{maplelatex}
\mapleinline{inert}{2d}{}{\[\displaystyle \xi_{{{\it y2}}} \left( {\it \sigma}\right) ={\it \xi 2}\\
\mbox{} \left( {\it y0},{\it y1},{\it y2},{\it y3} \right) \]}
\end{maplelatex}
\mapleresult
\begin{maplelatex}
\mapleinline{inert}{2d}{}{\[\displaystyle \xi_{{{\it y3}}} \left( {\it \sigma}\right) ={\it \xi 3}\\
\mbox{} \left( {\it y3} \right) \]}
\end{maplelatex}
\mapleresult
\begin{maplelatex}
\mapleinline{inert}{2d}{}{\[\displaystyle
\eta_{{{\it J0}}} \left( {\it \sigma}\right)=
{\it J1}\,{\frac {\partial }{\partial {\it y1}}}{\it \xi 0} \left( {\it y0},{\it y1},{\it y2},{\it y3}\\\mbox{} \right)
+{\it J2}\,{\frac {\partial }{\partial {\it y2}}}{\it \xi 0} \left( {\it y0},{\it y1},{\it y2},{\it y3}\\\mbox{} \right) \\\mbox{}
-{\it J0}\,{\frac {\partial }{\partial {\it y1}}}{\it \xi 1} \left( {\it y0},{\it y1},{\it y2},{\it y3}\\\mbox{} \right) \\\mbox{}
\]}
\end{maplelatex}
\mapleresult
\begin{maplelatex}
\mapleinline{inert}{2d}{}{\[\displaystyle
-{\it J0}\,{\frac {\partial }{\partial {\it y2}}}{\it \xi 2} \left( {\it y0},{\it y1},{\it y2},{\it y3}\\\mbox{} \right)
-C \left( {\it y3}\\\mbox{} \right) {\it J0}
+{\it b0} \left( {\it y0},{\it y1},{\it y2},{\it y3}\\\mbox{} \right)
\]}
\end{maplelatex}
\mapleresult
\begin{maplelatex}
\mapleinline{inert}{2d}{}{\[\displaystyle
\eta_{{{\it J1}}} \left( {\it \sigma}\right)=
{\it J0}\,{\frac {\partial }{\partial {\it y0}}}{\it \xi 1} \left( {\it y0},{\it y1},{\it y2},{\it y3}\\\mbox{} \right)
+{\it J2}\,{\frac {\partial }{\partial {\it y2}}}{\it \xi 1} \left( {\it y0},{\it y1},{\it y2},{\it y3}\\\mbox{} \right) \\\mbox{}
-{\it J1}\,{\frac {\partial }{\partial {\it y0}}}{\it \xi 0} \left( {\it y0},{\it y1},{\it y2},{\it y3}\\\mbox{} \right)
\]}
\end{maplelatex}
\mapleresult
\begin{maplelatex}
\mapleinline{inert}{2d}{}{\[\displaystyle
-{\it J1}\,{\frac {\partial }{\partial {\it y2}}}{\it \xi 2} \left( {\it y0},{\it y1},{\it y2},{\it y3}\\\mbox{} \right)
-C \left( {\it y3}\\\mbox{} \right) {\it J1}
+{\it b1} \left( {\it y0},{\it y1},{\it y2},{\it y3}\\\mbox{} \right) \\\mbox{}
\]}
\end{maplelatex}
\mapleresult
\begin{maplelatex}
\mapleinline{inert}{2d}{}{\[\displaystyle
\eta_{{{\it J2}}} \left( {\it \sigma}\right)=
{\it J0}\,{\frac {\partial }{\partial {\it y0}}}{\it \xi 2} \left( {\it y0},{\it y1},{\it y2},{\it y3}\\\mbox{} \right)
+{\it J1}\,{\frac {\partial }{\partial {\it y1}}}{\it \xi 2} \left( {\it y0},{\it y1},{\it y2},{\it y3}\\\mbox{} \right) \\\mbox{}
-{\it J2}\,{\frac {\partial }{\partial {\it y0}}}{\it \xi 0} \left( {\it y0},{\it y1},{\it y2},{\it y3}\\\mbox{} \right)
\]}
\end{maplelatex}
\mapleresult
\begin{maplelatex}
\mapleinline{inert}{2d}{}{\[\displaystyle
-{\it J2}\,{\frac {\partial }{\partial {\it y1}}}{\it \xi 1} \left( {\it y0},{\it y1},{\it y2},{\it y3}\\\mbox{} \right)
-C \left( {\it y3}\\\mbox{} \right) {\it J2}
+{\it b2} \left( {\it y0},{\it y1},{\it y2},{\it y3}\\\mbox{} \right) \\\mbox{}
\]}
\end{maplelatex}
\end{maplegroup}
%
%
\begin{maplegroup}
\begin{flushleft}
{\large Counter-checking above result if truly a solution --- and indeed it is:}
\end{flushleft}
\end{maplegroup}
\begin{maplegroup}
\begin{mapleinput}
\mapleinline{active}{1d}{rho:=(y0,y1,y2,y3):}{}
\end{mapleinput}
\begin{mapleinput}
\mapleinline{active}{1d}{xi[y0](sigma):=xi0(rho); xi[y1](sigma):=xi1(rho);
\hspace{0.43cm} xi[y2](sigma):=xi2(rho); xi[y3](sigma):=xi3(y3);
\hspace{0.43cm} eta[J0](sigma):=J1*(diff(xi0(rho),y1))+J2*(diff(xi0(rho),y2))
\hspace{0.43cm} -J0*(diff(xi1(rho),y1))-J0*(diff(xi2(rho),y2))-C(y3)*J0+b0(rho);
\hspace{0.43cm} eta[J1](sigma):=J0*(diff(xi1(rho),y0))+J2*(diff(xi1(rho),y2))
\hspace{0.43cm} -J1*(diff(xi0(rho),y0))-J1*(diff(xi2(rho),y2))-C(y3)*J1+b1(rho);
\hspace{0.43cm} eta[J2](sigma):=J0*(diff(xi2(rho),y0))+J1*(diff(xi2(rho),y1))
\hspace{0.43cm} -J2*(diff(xi0(rho),y0))-J2*(diff(xi1(rho),y1))-C(y3)*J2+b2(rho);
}{}
\end{mapleinput}
\mapleresult
\begin{maplelatex}
\mapleinline{inert}{2d}{}{\[\displaystyle
\xi_{{{\it y0}}} \left( {\it \sigma}\right)={\it \xi 0}\\\mbox{} \left(\rho \right),\;
\xi_{{{\it y1}}} \left( {\it \sigma}\right) ={\it \xi 1}\\\mbox{} \left( \rho \right),\;
\xi_{{{\it y2}}} \left( {\it \sigma}\right) ={\it \xi 2}\\\mbox{} \left( \rho \right),\;
\xi_{{{\it y3}}} \left( {\it \sigma}\right) ={\it \xi 3}\\\mbox{} \left( {\it y3} \right)
\]}
\end{maplelatex}
\mapleresult
\begin{maplelatex}
\mapleinline{inert}{2d}{}{\[\displaystyle
\eta_{{{\it J0}}} \left( {\it \sigma}\right)=
{\it J1}\,{\frac {\partial }{\partial {\it y1}}}{\it \xi 0} \left( \rho\\\mbox{} \right)
\!+\!{\it J2}\,{\frac {\partial }{\partial {\it y2}}}{\it \xi 0} \left(\rho\\\mbox{} \right) \\\mbox{}
\!-\!{\it J0}\,{\frac {\partial }{\partial {\it y1}}}{\it \xi 1} \left( \rho\\\mbox{} \right) \\\mbox{}
\!-\!{\it J0}\,{\frac {\partial }{\partial {\it y2}}}{\it \xi 2} \left( \rho\\\mbox{} \right)
\!-\!C \left( {\it y3}\\\mbox{} \right) {\it J0}
\!+\!{\it b0} \left(\rho\\\mbox{} \right)
\]}
\end{maplelatex}
\mapleresult
\begin{maplelatex}
\mapleinline{inert}{2d}{}{\[\displaystyle
\eta_{{{\it J1}}} \left( {\it \sigma}\right)=
{\it J0}\,{\frac {\partial }{\partial {\it y0}}}{\it \xi 1} \left( \rho\\\mbox{} \right)
\!+\!{\it J2}\,{\frac {\partial }{\partial {\it y2}}}{\it \xi 1} \left( \rho\\\mbox{} \right) \\\mbox{}
\!-\!{\it J1}\,{\frac {\partial }{\partial {\it y0}}}{\it \xi 0} \left( \rho\\\mbox{} \right)
\!-\!{\it J1}\,{\frac {\partial }{\partial {\it y2}}}{\it \xi 2} \left( \rho\\\mbox{} \right)
\!-\!C \left( {\it y3}\\\mbox{} \right) {\it J1}
\!+\!{\it b1} \left( \rho\\\mbox{} \right) \\\mbox{}
\]}
\end{maplelatex}
\mapleresult
\begin{maplelatex}
\mapleinline{inert}{2d}{}{\[\displaystyle
\eta_{{{\it J2}}} \left( {\it \sigma}\right)=
{\it J0}\,{\frac {\partial }{\partial {\it y0}}}{\it \xi 2} \left( \rho\\\mbox{} \right)
\!+\!{\it J1}\,{\frac {\partial }{\partial {\it y1}}}{\it \xi 2} \left( \rho\\\mbox{} \right) \\\mbox{}
\!-\!{\it J2}\,{\frac {\partial }{\partial {\it y0}}}{\it \xi 0} \left( \rho\\\mbox{} \right)
\!-\!{\it J2}\,{\frac {\partial }{\partial {\it y1}}}{\it \xi 1} \left( \rho\\\mbox{} \right)
\!-\!C \left( {\it y3}\\\mbox{} \right) {\it J2}
\!+\!{\it b2} \left( \rho\\\mbox{} \right) \\\mbox{}
\]}
\end{maplelatex}
\vspace{1em}
\begin{mapleinput}
\mapleinline{active}{1d}{simplify(detsys[1]);
}{}
\end{mapleinput}
\mapleresult
\begin{maplelatex}
\mapleinline{inert}{2d}{}{\[\displaystyle
\bigg[0,0,0,0,0,0,0,0,0,0,0,0,0,0,0,0,0,0,0,0,0,0,0,0,0,0,0,0,0,0,0,
\]}
\end{maplelatex}
\mapleresult
\begin{maplelatex}
\mapleinline{inert}{2d}{}{\[\displaystyle
\hspace{0.1cm} 0,0,0,0,0,0,0,0,0,0,0,
{\frac {\partial }{\partial {\it y0}}}{\it b0} \left( \rho \right)
+{\frac {\partial }{\partial {\it y1}}}{\it b1} \left( \rho \right)
+{\frac {\partial }{\partial {\it y2}}}{\it b2} \left( \rho\right) \\\mbox{} ,0,0 \bigg]
\]}
\end{maplelatex}
\end{maplegroup}

%% file: V2.tex
\begin{maplegroup}
\begin{flushleft}
{\large Header:}
\end{flushleft}
\end{maplegroup}
\begin{maplegroup}
\begin{mapleinput}
\mapleinline{active}{1d}{restart: read "Desolv-V5R5.mpl": with(desolv):}{}
\end{mapleinput}
\mapleresult
\begin{maplelatex}
\mapleinline{inert}{2d}{`DESOLVII_V5R5 (March-2011)(c) by Dr. K.
T. Vu, Dr. J. Carminati and
Miss. G. Jefferson`;}{%
\maplemultiline{ \mathit{\phantom{xxxxxxxx} DESOLVII\_V5R5\ (March-2011)(c)} \\
\mathit{by\ Dr.\ K.\ T.\ Vu,\ Dr.\ J.\ Carminati\ and\ Miss.\ G.\
Jefferson} }}
\end{maplelatex}
\end{maplegroup}
\begin{maplegroup}
\begin{flushleft}
{\large Definitions of variables and local equation:}
\end{flushleft}
\end{maplegroup}
\begin{maplegroup}
\begin{mapleinput}
\mapleinline{active}{1d}{alias(sigma=(y0,y1,y2,y3,J0,J1,J2)): X:=(y0,y1,y2,y3):
}{}
\end{mapleinput}
\end{maplegroup}
\begin{maplegroup}
\begin{mapleinput}
\mapleinline{active}{1d}{eqn:=diff(J0(X),y0)+diff(J1(X),y1)+diff(J2(X),y2)=0:
}{}
\end{mapleinput}
\end{maplegroup}
\begin{maplegroup}
\begin{flushleft}
{\large Symmetry Algorithm:}\\[0.75em]
\textit{{\large Size of the determining system:}}
\end{flushleft}
\end{maplegroup}
\begin{maplegroup}
\begin{mapleinput}
\mapleinline{active}{1d}{detsys:=gendef([eqn],[J0,J1,J2],[y0,y1,y2,y3]): nops(detsys[1]);
}{}
\end{mapleinput}
\mapleresult
\begin{maplelatex}
\mapleinline{inert}{2d}{24}{\[\displaystyle 24\]}
\end{maplelatex}
\end{maplegroup}
\vspace{-0.75em}
\begin{maplegroup}
\begin{flushleft}
\textit{{\large Solving the determining system:}}
\end{flushleft}
\end{maplegroup}
\begin{maplegroup}
\begin{mapleinput}
\mapleinline{active}{1d}{sym:=pdesolv(op(detsys));
}{}
\end{mapleinput}
\mapleresult
\begin{maplelatex}
\mapleinline{inert}{2d}{}{\[\displaystyle
{\it sym}\, := \,\bigg[\bigg[-{\frac {\partial }{\partial {\it y1}}}{\it F\_6} \left( {\it y0},{\it y1},{\it y2},{\it y3}\\ \mbox{} \right)
-{\frac {\partial }{\partial {\it y2}}}{\it F\_9} \left( {\it y0},{\it y1},{\it y2},{\it y3}\\
\mbox{} \right) -{\it F\_26} \left( {\it y0},{\it y1},{\it y2},{\it y3}\\ \mbox{} \right),
\]}
\end{maplelatex}
\mapleresult
\begin{maplelatex}
\mapleinline{inert}{2d}{}{\[\displaystyle
{\frac {\partial }{\partial {\it y0}}}{\it F\_23} \left( {\it y0},{\it y1},{\it y2},{\it y3}\\
\mbox{} \right)
+{\frac {\partial }{\partial {\it y1}}}{\it F\_24} \left( {\it y0},{\it y1},{\it y2},{\it y3}\\
\mbox{} \right)
+{\frac {\partial }{\partial {\it y2}}}{\it F\_25} \left( {\it y0},{\it y1},{\it y2},{\it y3}\\
\mbox{} \right) \\
\mbox{} \bigg],\,[\,],
\]}
\end{maplelatex}
\mapleresult
\begin{maplelatex}
\mapleinline{inert}{2d}{}{\[\displaystyle
\bigg[\xi_{{{\it y0}}} \left( {\it \sigma}\right)\!=\!{\it F\_3}\\
\mbox{} \left( {\it y0},{\it y1},{\it y2},{\it y3} \right),
\xi_{{{\it y1}}} \left( {\it \sigma}\right)\!=\!{\it F\_6} \left( {\it y0},{\it y1},{\it y2},{\it y3} \right),
\xi_{{{\it y2}}} \left( {\it \sigma}\right)\\
\mbox{}\!=\!{\it F\_9} \left( {\it y0},{\it y1},{\it y2},{\it y3} \right),
\]}
\end{maplelatex}
\mapleresult
\begin{maplelatex}
\mapleinline{inert}{2d}{}{\[\displaystyle
\xi_{{{\it y3}}} \left( {\it \sigma}\right)\!=\!{\it F\_15} \left( {\it y3} \right),
\]}
\end{maplelatex}
\mapleresult
\begin{maplelatex}
\mapleinline{inert}{2d}{}{\[\displaystyle
\eta_{{{\it J0}}} \left( {\it \sigma}\right)\!=\!
{\it F\_26} \left( {\it y0},{\it y1},{\it y2},{\it y3} \right){\it J0}
\!+\!{\it J1}\,{\frac {\partial }{\partial {\it y1}}}{\it F\_3}\\
\mbox{} \left( {\it y0},{\it y1},{\it y2},{\it y3} \right)
\]}
\end{maplelatex}
\mapleresult
\vspace{-0.5em}
\begin{maplelatex}
\mapleinline{inert}{2d}{}{\[\displaystyle
\hspace{1.5cm}+{\it J2}\,{\frac {\partial }{\partial {\it y2}}}{\it F\_3}\\
\mbox{} \left( {\it y0},{\it y1},{\it y2},{\it y3} \right)\!+\!{\it F\_23} \left( {\it y0},{\it y1},{\it y2},{\it y3} \right) \\
\mbox{},
\]}
\end{maplelatex}
\vspace{-0.5em}
\mapleresult
\begin{maplelatex}
\mapleinline{inert}{2d}{}{\[\displaystyle
\eta_{{{\it J1}}} \left( {\it \sigma}\right)\!=\!
{\it J0}\,{\frac {\partial }{\partial {\it y0}}}{\it F\_6} \left( {\it y0},{\it y1},{\it y2},{\it y3} \right) \\
\mbox{}\!+\!{\it F\_26} \left( {\it y0},{\it y1},{\it y2},{\it y3} \right) {\it J1}
\!+\!{\it J2}\,{\frac {\partial }{\partial {\it y2}}}{\it F\_6} \left( {\it y0},{\it y1},{\it y2},{\it y3} \right)
\]}
\end{maplelatex}
\mapleresult
\vspace{-0.25em}
\begin{maplelatex}
\mapleinline{inert}{2d}{}{\[\displaystyle
\hspace{1.5cm}+{\it J1}\,{\frac {\partial }{\partial {\it y1}}}{\it F\_6} \left( {\it y0},{\it y1},{\it y2},{\it y3} \right)
\!-\!{\it J1}\,{\frac {\partial }{\partial {\it y0}}}{\it F\_3}\\
\mbox{} \left( {\it y0},{\it y1},{\it y2},{\it y3} \right)\!+\!{\it F\_24} \left( {\it y0},{\it y1},{\it y2},{\it y3} \right) ,
\]}
\end{maplelatex}
\mapleresult
\begin{maplelatex}
\mapleinline{inert}{2d}{}{\[\displaystyle
\eta_{{{\it J2}}} \left( {\it \sigma}\right) \\
\mbox{}\!=\!
{\it J0}\,{\frac {\partial }{\partial {\it y0}}}{\it F\_9} \left( {\it y0},{\it y1},{\it y2},{\it y3} \right)
\!+\!\,{\it F\_26} \left( {\it y0},{\it y1},{\it y2},{\it y3} \right){\it J2}
\!+\!{\it J2}\,{\frac {\partial }{\partial {\it y2}}}{\it F\_9} \left( {\it y0},{\it y1},{\it y2},{\it y3} \right)
\]}
\end{maplelatex}
\mapleresult
\begin{maplelatex}
\mapleinline{inert}{2d}{}{\[\displaystyle
\hspace{1.5cm}+{\it J1}\,{\frac {\partial }{\partial {\it y1}}}{\it F\_9} \left( {\it y0},{\it y1},{\it y2},{\it y3} \right)
\!-\!{\it J2}\,{\frac {\partial }{\partial {\it y0}}}{\it F\_3}\\
\mbox{} \left( {\it y0},{\it y1},{\it y2},{\it y3} \right)
\!+\!{\it F\_25} \left( {\it y0},{\it y1},{\it y2},{\it y3} \right)\\
\mbox{}\bigg],
\]}
\end{maplelatex}
\mapleresult
\begin{maplelatex}
\mapleinline{inert}{2d}{}{\[\displaystyle
\bigg[{\it F\_3} \left( {\it y0},{\it y1},{\it y2},{\it y3}\\
\mbox{} \right) ,{\it F\_6} \left( {\it y0},{\it y1},{\it y2},{\it y3}\\
\mbox{} \right) ,{\it F\_9} \left( {\it y0},{\it y1},{\it y2},{\it y3}\\
\mbox{} \right) ,{\it F\_23} \left( {\it y0},{\it y1},{\it y2},{\it y3}\\
\mbox{} \right), \\
\mbox{}
\]}
\end{maplelatex}
\mapleresult
\begin{maplelatex}
\mapleinline{inert}{2d}{}{\[\displaystyle
{\it F\_24} \left( {\it y0},{\it y1},{\it y2},{\it y3}\\
\mbox{} \right) ,{\it F\_25} \left( {\it y0},{\it y1},{\it y2},{\it y3}\\
\mbox{} \right) ,{\it F\_26} \left( {\it y0},{\it y1},{\it y2},{\it y3}\\
\mbox{} \right) \\
\mbox{},
{\it F\_15} \left( {\it y3}\\
\mbox{} \right)\bigg]\bigg]
\]}
\end{maplelatex}
\end{maplegroup}
\begin{maplegroup}
\begin{flushleft}
{\large This is exactly the same solution as obtained in Version No.$\,$1.
Just perform the following renaming,}
\end{flushleft}
\end{maplegroup}
\vspace{-1em}
\begin{maplegroup}
\mapleresult
\begin{maplelatex}
\mapleinline{inert}{2d}{}{\[\displaystyle
{\it F\_3}\rightarrow {\it F\_27},\;
{\it F\_6}\rightarrow {\it F\_15},\;
{\it F\_9}\rightarrow {\it F\_21},\;
{\it F\_15}\rightarrow {\it F\_9},\;
\]}
\end{maplelatex}
\mapleresult
\begin{maplelatex}
\mapleinline{inert}{2d}{}{\[\displaystyle
{\it F\_23}\rightarrow {\it F\_44},\;
{\it F\_24}\rightarrow {\it F\_45},\;
{\it F\_25}\rightarrow {\it F\_46},\;
{\it F\_26}\rightarrow {\it F\_47}
\]}
\end{maplelatex}
\end{maplegroup}